\documentclass[12pt]{iopart}
\pdfoutput=1

\usepackage{iopams}
\usepackage{graphicx}

\begin{document}

\title[Improvement of two-way continuous-variable quantum key distribution]{Improvement of two-way continuous-variable quantum key distribution using optical amplifiers}

\author{Yi-Chen Zhang$^1$, Zhengyu Li$^2$, Christian Weedbrook$^3$, Song Yu$^1$, Wanyi Gu$^1$, Maozhu Sun$^2$, Xiang Peng$^2$, Hong Guo$^2$}

\address{$^1$The State Key Laboratory of Information Photonics and Optical Communications, Beijing University of Posts and Telecommunications, Beijing 100876, China}

\address{$^2$The State Key Laboratory of Advanced Optical Communication Systems and Networks, School of Electronics Engineering and Computer Science, Peking University, Beijing 100871, China}

\address{$^3$Department of Physics, University of Toronto, Toronto, M5S 3G4, Canada}

\ead{yusong@bupt.edu.cn and hongguo@pku.edu.cn}

\begin{abstract}
The imperfections of a receiver's detector affect the performance of two-way continuous-variable quantum key distribution protocols and are difficult to adjust in practical situations. We propose a method to improve the performance of two-way continuous-variable quantum key distribution by adding a parameter-adjustable optical amplifier at the receiver. A security analysis is derived against a two-mode collective entangling cloner attack. Our simulations show that the proposed method can improve the performance of protocols as long as the inherent noise of the amplifier is lower than a critical value, defined as the tolerable amplifier noise. Furthermore, the optimal performance can approach the scenario where a perfect detector is used.
\end{abstract}

\submitto{\JPB}
\maketitle

\section{Introduction}
Quantum key distribution (QKD)~\cite{Gisin_RevModPhys_2002,Scarani_RevModPhys_2009} is one of the most practical applications in the field of quantum information. Its goal is to establish a secure key between two legitimate partners, usually called Alice and Bob. Continuous-variable quantum key distribution (CV-QKD)~\cite{Weedbrook_RevModPhys2012} has attracted much attention in the past few years~\cite{Scarani_RevModPhys_2009,Weedbrook_RevModPhys2012,Jouguet_nature_2013} mainly because it only uses standard telecom components. A CV-QKD protocol based on coherent states~\cite{Grosshans_PhysRevLett_2002,Weedbrook_PhysRevLett_2004} with Gaussian modulation has been experimentally demonstrated~\cite{grosshans_nature_2003,Lance_PRL_2005,Lodewyck_PhysRevA_2007,Jouguet_nature_2013} and has been shown to be secure against arbitrary collective attacks~\cite{Grosshans_PhysRevLett_2005,Navascues_PhysRevLett_2005}. Such an attack is the most optimal in the asymptotical limit~\cite{Renner_PhysRevLett_2009} and is also used in the finite-size regime~\cite{Furrer_PhysRevLett_2012,Leverrier_PhysRevLett_2013}.

To enhance the tolerable excess noise of CV-QKD, compared to the typical one-way schemes, the two-way CV-QKD was proposed~\cite{pirandola_nature_2008}. Recently, a more feasible two-way CV-QKD protocol was proposed by replacing Alice's displacement operation with a beam splitter and inserting thermal noise into it. This leads to a protocol that is easier to analyze when considering channel estimation~\cite{sunmaozhu_WorldScientific_2012}.

In practice, the detector's imperfections, mainly characterized by the detection efficiency and electronic noise, will affect the performance of two-way CV-QKD protocols and are hard to adjust in an experiment~\cite{Jouguet_nature_2013,Lodewyck_PhysRevA_2007}. In this paper, we insert an optical amplifier before Bob's detection by which the receiver's efficiency and noise can be optimized to improve the performance of two-way CV-QKD protocols using reverse reconciliation. Previously, similar method had only been analyzed for the case of one-way~\cite{Fossier_JPB_2009} and one-way four-state~\cite{ZhangHeng_PhysRevA_2012} reverse reconciliation schemes. Using numerical simulations, we propose a specific two-mode collective entangling cloner attack as Eve's action.

The paper is organized as follows. In Sec.~\ref{sec:2}, we review the two-way CV-QKD protocols with imperfect detectors and describe the extended model of the detector. In Sec.~\ref{sec:3}, we first analyze the different optical amplifier models and their suitable scopes of application. Then we derive security bounds of the protocols with different optical amplifiers. Finally, the simulation results under collective entangling cloner attack are provided to compare the performances of the protocols with and without the amplifiers. Our conclusions are drawn in Sec.~\ref{sec:4}.

\begin{figure*}[t]
\centering
\includegraphics[height=2in]{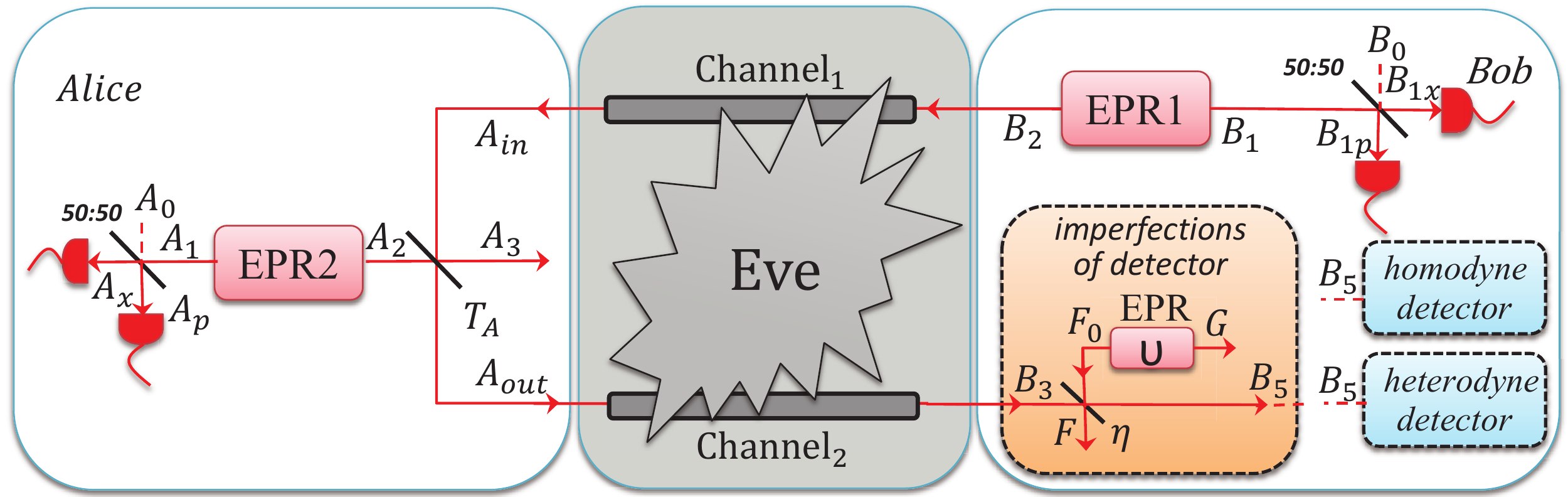}
\caption{Entanglement-based scheme of Gaussian-modulated two-way CV-QKD protocols with imperfect homodyne or heterodyne detection where the quantum channel is fully controlled by Eve. However, Eve has no access to the apparatuses in Alice's and Bob's stations.
}\label{fig1}
\end{figure*}

\section{\label{sec:2}Security analysis of two-way CV-QKD protocols with imperfect detector}
In the following, we first review the basic notions of the entanglement-based model related to the Gaussian-modulated two-way CV-QKD protocols with imperfect detection~\cite{sunmaozhu_WorldScientific_2012,sunmaozhu_JPB_2012}. The entanglement-based model with imperfect detectors is illustrated in~Fig.~\ref{fig1} and can be described as follows:
\\
\\{Step 1:} Bob initially prepares an EPR pair (EPR1 with variance $V_{B}$, where the shot noise variance is normalized to $1$), keeps one mode $B_{1}$ and sends the other mode $B_{2}$ to Alice through the channel where Eve may perform her attack.
\\
\\{Step 2:} Alice prepares another EPR pair (EPR2 with variance $V_{A}$). She keeps mode $A_{1}$ and measures it using heterodyne detection to get the variables {$x_{A_{x}}$,$p_{A_{p}}$}. The modes $A_0$ and $B_0$ represent vacuum state. She then couples mode $A_{2}$ and the received mode $A_{in}$ from Bob with a beam splitter (transmittance: $T_{A} \in [0,1]$). Alice then sends mode $A_{out}$ back to Bob, and measures another mode $A_{3}$ with homodyne detection for parameter estimation~\cite{sunmaozhu_WorldScientific_2012}.
\\
\\{Step 3:} Bob measures his original mode $B_{1}$ using heterodyne detection to get the variables $x_{B_{1x}}$ and $p_{B_{1p}}$. He also measures the received mode $B_{5}$ with homodyne detection to get $x_{B_{5}}$ or with heterodyne detection to get $x_{B_{5x}}$ and $p_{B_{5p}}$. The detector's inefficiency is modelled by a beam splitter with transmittance $\eta$, while its electronic noise $\upsilon_{el}$ is modelled by a thermal state ${\rho _{{F_0}}}$ with variance $\upsilon$ \cite{Jouguet_nature_2013}.
\\
\\{Step 4:} When Bob uses homodyne detection, he uses ${x_{B_x}} = {x_{B_5}} - {k} x_{B_{1x}}$ (${p_{B_p}} = {p_{B_5}} - {k} p_{B_{1p}}$) to construct the estimator to Alice's corresponding variable $x_{{A_x}}$ ($p_{{A_p}}$), where ${k}$ is the parameter used to optimize Bob's estimator of Alice¡¯s corresponding value. When Bob uses heterodyne detection, he uses a similar way to construct the estimators (${x_{B_x}} = {x_{B_{5x}}} - {k} x_{B_{1x}}$, ${p_{B_p}} = {p_{B_{5p}}} - {k} p_{B_{1p}}$) to ($x_{{A_x}}$ , $p_{{A_p}}$) at the same time. Then Alice and Bob proceed with classical data postprocessing namely reconciliation and privacy amplification. In this paper we use reverse reconciliation~\cite{grosshans_nature_2003}.
\\

\begin{figure}[t]
\centering
\includegraphics[height=2in]{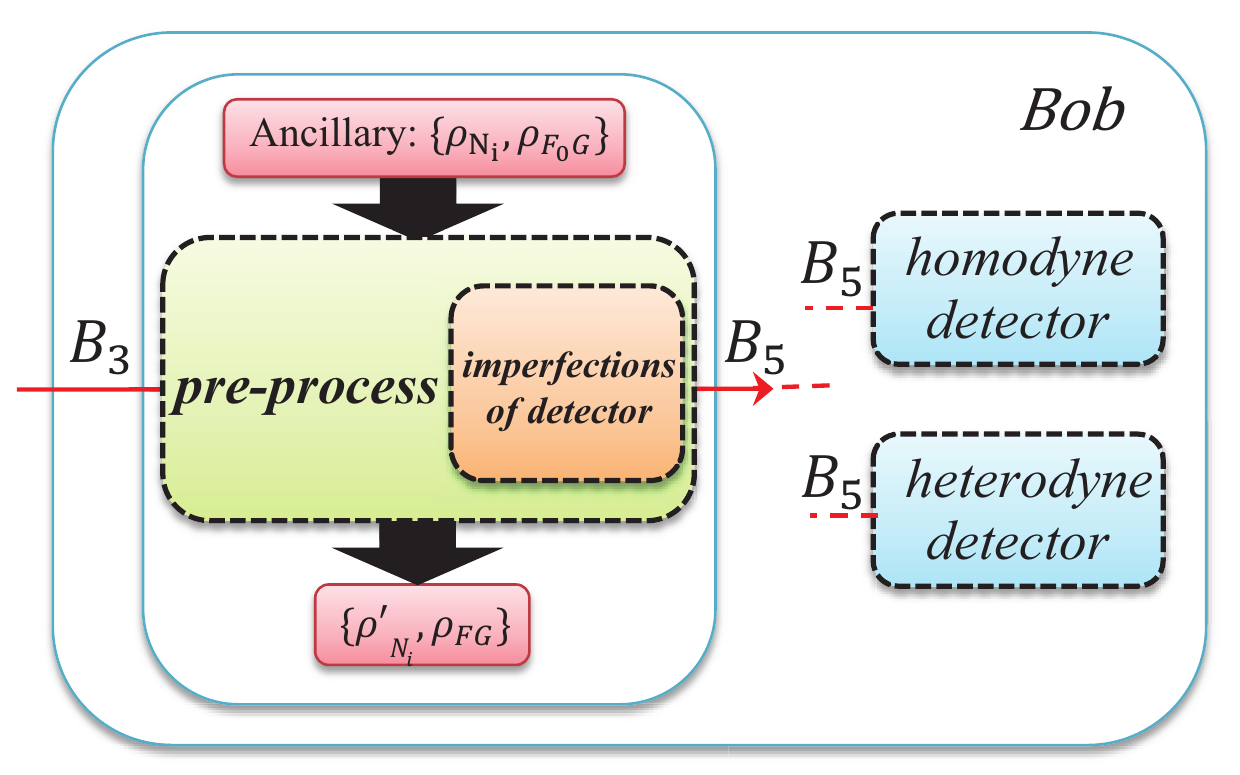}
\caption{The receiver model which consists of a \emph{pre-process} phase and the imperfections of detector with the ancillas $\rho_{N_{i}}$ and $\rho_{F_{0}G}$, respectively.
}\label{fig2}
\end{figure}

In order for Alice and Bob to perform their measurement, Bob initially sends a local oscillator with the signal beam to Alice, then Alice couples it with the local oscillator from her and sends back to Bob with the signal beam. The variance $\upsilon$  of the thermal state ${\rho _{{F_0}}}$ is chosen to obtain the appropriate expression for each detection, in the following way: for homodyne detection, $\upsilon  = 1 + {{{\upsilon _{el}}} \mathord{\left/
 {\vphantom {{{\upsilon _{el}}} {\left( {1 - \eta } \right)}}} \right.
 \kern-\nulldelimiterspace} {\left( {1 - \eta } \right)}}$, and for heterodyne detection, $\upsilon  = 1 + {{2{\upsilon _{el}}} \mathord{\left/
 {\vphantom {{2{\upsilon _{el}}} {\left( {1 - \eta } \right)}}} \right.
 \kern-\nulldelimiterspace} {\left( {1 - \eta } \right)}}$~\cite{Lodewyck_PhysRevA_2007}.
Adjusting the efficiency $\eta$ and the variance $\upsilon$, we can optimize the performance of the protocols. For instance, the performance of two-way CV-QKD can be improved by adding noise in homodyne detection \cite{sunmaozhu_JPB_2012}. Unfortunately, for a practical detector, the detection efficiency $\eta$ and electronic noise $\upsilon$ are fixed between $B_3$ and $B_5$ (see Fig.~\ref{fig1}), and generally speaking they are not the optimal choice. To improve the performance of the protocols, we can insert an adjustable operation with ancilla $\rho_{N_{i}}$ before detection, noted as the \emph{pre-process} (see Fig.~\ref{fig2}). Therefore the \emph{pre-process} phase and the imperfections of the detector constitute a new receiver, whose efficiency and noise can be optimized to improve the performances of the two-way schemes.

\section{\label{sec:3}Improvement of two-way CV-QKD protocols with optical amplifiers}
In this section, we present two kinds of optical amplifier models \cite{Caves_PhysRevD_1982,Fossier_JPB_2009,ZhangHeng_PhysRevA_2012}: a perfect phase-sensitive amplifier (PSA) and a practical phase-insensitive amplifier (PIA), as the \emph{pre-process} to improve the performances of homodyne and heterodyne detection two-way CV-QKD protocols, respectively. Simulation results against a two-mode collective entangling cloner attack are provided to compare the performances of the protocols with and without the amplifiers.

\subsection{Homodyne detection with PSA}
A PSA is a degenerate optical parametric amplifier, which permits noiseless amplification of a chosen quadrature ($\hat{x}$ or $\hat{p}$)~\cite{Caves_PhysRevD_1982}. Its mathematical model can be described by the transformation matrix $Y^{PSA}$
\begin{equation}
{\left[ {\begin{array}{*{20}{c}}
   {\hat x}  \\
   {\hat p}  \\
\end{array}} \right]_{out}} = \left[ {\begin{array}{*{20}{c}}
   {\sqrt g } & 0  \\
   0 & {{1 \mathord{\left/
 {\vphantom {1 {\sqrt g }}} \right.
 \kern-\nulldelimiterspace} {\sqrt g }}}  \\
\end{array}} \right] \cdot {\left[ {\begin{array}{*{20}{c}}
   {\hat x}  \\
   {\hat p}  \\
\end{array}} \right]_{in}} \buildrel\textstyle\over= {Y^{PSA}} \cdot {\left[ {\begin{array}{*{20}{c}}
   {\hat x}  \\
   {\hat p}  \\
\end{array}} \right]_{in}},
\end{equation}
where $g > 1$ is the gain of the optical amplifier, $g = 1$ means the optical amplifier does nothing.

\begin{figure}[t]
\centering
\includegraphics[height=2in]{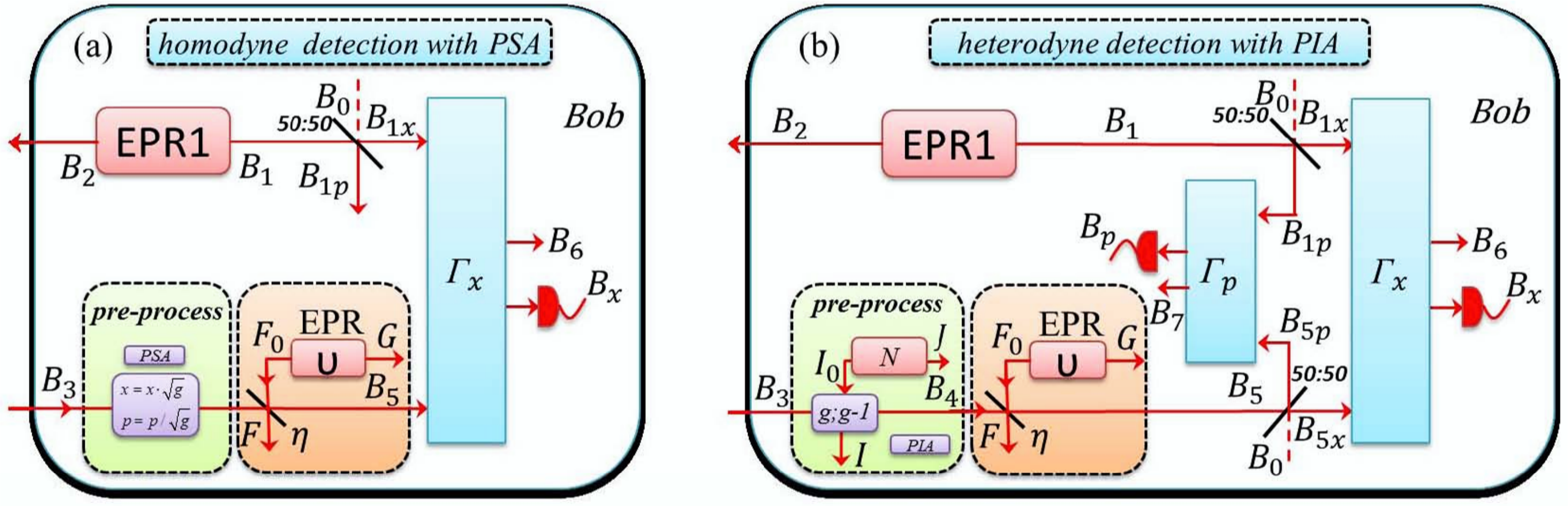}
\caption{(a) The scheme for homodyne detection two-way CV-QKD protocol when adding a phase sensitive amplifier at the receiver, where the CNOT gate ${\Gamma _x}$ \cite{G_Patron_PhD_2007,Yoshikawa_PhysRevLett_2008,Nielsen_QCQI} refers to the postprocessing stage in reverse reconciliation. (b) The scheme for heterodyne detection two-way CV-QKD protocol when adding a phase insensitive amplifier at the receiver.
}\label{fig3}
\end{figure}

We now derive security bound of the two-way CV-QKD protocol with homodyne detection adding a PSA before detection at the receiver (see Fig.~\ref{fig3} (a)). When Alice and Bob use reverse reconciliation, the secret key rate is given by

\begin{equation}
K=\beta I(a:b)-S(b:E),
\end{equation}
where $\beta \in[0,1]$ is the reconciliation efficiency, $I(a:b)$ is the classical mutual information between Alice and Bob, and $S(b:E)$ is the quantum mutual information between Bob and Eve. The classical mutual information between Alice and Bob can be written as
\begin{equation}
I(a:b)=\frac{1}{2}\log V_{A_{x}}-\frac{1}{2}\log V_{A_{x}|B_x},
\end{equation}
where ${V_{{A_x}}} = \frac{1}{2}\left( {{V_A} + 1} \right)$, ${V_{{A_x}|{B_x}}}$ is the variance of mode $A_{x}$ conditioned on Bob's data, and Bob using ${x_{{B_x}}} = {x_{{B_5}}} - k  {x_{{B_{1x}}}}$ as his final data. The state Bob gets after total channels is

\begin{equation}
{\hat B_5} = \sqrt {\eta g{T_A}{T_1}{T_2}} {\hat B_2} + \sqrt {{1 - {T_A}}{T_2}\eta g} {\hat A_2} + \sqrt {1 - \eta } {\hat F_0} + \hat E,
\end{equation}
where $\hat E$ represents the total excess noise introduced by Eve. Bob uses ${x_{{B_5}}}$ and ${x_{{B_{1x}}}}$ to construct his estimator of Alice's variable ${x_{{A_1}}}$, ${x_{{B_x}}} = {x_{{B_5}}} - k{x_{{B_{1x}}}}$, where ${x_{{B_5}}}$, ${x_{{B_{1x}}}}$ and ${x_{{A_1}}}$ are the measurement results of the modes $\hat B_5$, $\hat B_{1x}$ and $\hat A_1$. To make Bob's estimator as precise as possible, he chooses a value of $k$ to reduce the interference from $\hat B_2$, specifically minimizing the variance of ${x_{{B_{x}}}}$

\begin{equation}
{k} = \sqrt {2\eta g{T_A}{T_1}{T_2} \Big(\frac{{{V_B} - 1}}{{{V_B} + 1}}\Big)} .
\end{equation}

The maximum information available to Eve on Bob's raw key is bounded by the Holevo bound~\cite{Nielsen_QCQI}

\begin{eqnarray}
{ S(b:E) \le \chi _{BE}} = S({\rho _E}) - \int {p({x_{{B_x}}})} S(\rho _E^{{x_{{B_x}}}})d{x_{{B_x}}},
\end{eqnarray}
where ${p\left( {{x_{{{B_x}}}}} \right)}$ is the probability density function of the measurement output, ${\rho _E^{{x_{B_x}}}}$ is the eavesdropper's state conditioned on Bob's measurement result ${x_{B_x}}$, and $S(\rho)$ is the von Neumann entropy of the quantum state $\rho $.

To calculate ${\chi _{BE}}$ we first have $S\left( {{\rho _{{A_1}{A_3}{B_1}{B_3}}}} \right) = S\left( {{\rho _E}} \right)$ since Eve can purify Alice and Bob's system ${A_1}{A_3}{B_1}{B_3}$~\cite{Text}. Second, after Bob's projective measurement resulting in $x_{B_x}$, the system ${A_1}{A_3}{B_{1p}}{B_6}EFG$ is pure, so that $S( {\rho _{{A_0}{A_3}{B_{1p}}{B_6}FG}^{{x_{{B_x}}}}}) = S( {\rho _E^{{x_{{B_x}}}}})$, where $S( {\rho _{{A_1}{A_3}{B_{1p}}{B_6}FG}^{{x_{{B_x}}}}})$ is independent of $x_{B_x}$ for protocols applying Gaussian modulation of Gaussian states. Thus, ${\chi _{BE}}$ becomes
\begin{equation}
{\chi _{BE}} = S({\rho _{{A_1}{A_3}{B_1}{B_3}}}) - S(\rho _{{A_1}{A_3}{B_{1p}}{B_6}FG}^{{x_{{B_x}}}}).
\end{equation}

The entropies $S({\rho _{{A_1}{A_3}{B_1}{B_3}}})$ and $S( {\rho _{{A_1}{A_3}{B_{1p}}{B_6}FG}^{{x_{B_x}}}})$ can be calculated using the covariance matrices ${{\gamma _{{A_1}{A_3}{B_1}{B_3}}}}$ characterizing the state ${{\rho _{{A_1}{A_3}{B_1}{B_3}}}}$ and ${\gamma _{{A_1}{A_3}{B_{1p}}{B_6}FG}^{{x_{{B_x}}}}}$ characterizing the state ${\rho _{{A_1}{A_3}{B_{1p}}{B_6}FG}^{{x_{{B_x}}}}}$. So the expression for ${\chi _{BE}}$ can be further simplified as follows
\begin{equation}\label{eq}
{\chi _{BE}} = \sum\limits_{i = 1}^4 {G\Big(\frac{{{\lambda _i} - 1}}{2}\Big)}  - \sum\limits_{i = 5}^{10} {G\Big(\frac{{{\lambda _i} - 1}}{2}\Big)},
\end{equation}
where $G(x) = (x + 1)\log_2 (x + 1) - x\log_2 x$, ${\lambda _{1 - 4}}$ are the symplectic eigenvalues of the covariance matrix ${{\gamma _{{A_1}{A_3}{B_1}{B_3}}}}$ and ${\lambda _{5 - 10}}$ are the symplectic eigenvalues of the covariance matrix ${\gamma _{{A_1}{A_3}{B_{1p}}{B_6}FG}^{x_{B_x}}}$. After Alice and Bob measure the mode $A_1$ with heterodyne detection and measure the modes $A_3$, $B_{1p}$, $B_{x}$ and $B_6$ with homodyne detection in Fig.~\ref{fig1} and Fig.~\ref{fig3} (a), we can get the covariance matrices ${{\gamma _{{A_1}{A_3}{B_1}{B_3}}}}$ and ${\gamma _{{A_1}{A_3}{B_{1p}}{B_6}FG}^{x_{B_x}}}$ in experiment. However, in numerical simulation, we could not have the measurement results and we need the specific description of Eve's attack~\cite{sunmaozhu_WorldScientific_2012, sunmaozhu_JPB_2012} to help us to obtain the final data to calculate the covariance matrices ${{\gamma _{{A_1}{A_3}{B_1}{B_3}}}}$ and ${\gamma _{{A_1}{A_3}{B_{1p}}{B_6}FG}^{x_{B_x}}}$, which we will describe in the later simulation and discussion part in detail.

\subsection{Heterodyne detection with PIA}
A PIA is a nondegenerate optical parametric amplifier, which amplifies both quadratures~\cite{Caves_PhysRevD_1982}. However, the amplification process is associated with a fundamental excess noise. As illustrated in Fig.~\ref{fig3} (b), its mathematical model can be described by a noiseless amplifier whose transformation matrix is $Y^{PIA}$ and an EPR state of variance $N$, one-half of which is entering the amplifier's second input port
\begin{equation}
\left[ {\begin{array}{*{20}{c}}
   {{{\hat B}_4}}  \\
   {\hat I}  \\
\end{array}} \right] = \left[ {\begin{array}{*{20}{c}}
   {\sqrt g \cdot{{\rm{I}}_2}} & {\sqrt {g - 1} \cdot{\sigma _z}}  \\
   {\sqrt {g - 1} \cdot{\sigma _z}} & {\sqrt g \cdot{{\rm{I}}_2}}  \\
\end{array}} \right] \cdot \left[ {\begin{array}{*{20}{c}}
   {{{\hat B}_3}}  \\
   {{{\hat I}_0}}  \\
\end{array}} \right] \buildrel\textstyle\over= {Y^{PIA}} \cdot \left[ {\begin{array}{*{20}{c}}
   {{{\hat B}_3}}  \\
   {{{\hat I}_0}}  \\
\end{array}} \right].
\end{equation}
The EPR state of variance $N$ is used to represent the inherent noise of the amplifier, and its covariance matrix is given by
\begin{equation}
{\gamma _{{I_0}J}} = \left[ {\begin{array}{*{20}{c}}
   {N \cdot {{\rm{I}}_2}} & {\sqrt {{N^2} - 1}  \cdot {\sigma _z}}  \\
   {\sqrt {{N^2} - 1}  \cdot {\sigma _z}} & {N \cdot {{\rm{I}}_2}}  \\
\end{array}} \right].
\end{equation}

We can now derive security bound of the two-way CV-QKD protocol with heterodyne detection when adding a PIA at the receiver's device. For the heterodyne detection, Bob uses ${x_{B_x}} = {x_{B_{5x}}} - {k} {x_{B_{1x}}}$ and ${p_{B_p}} = {p_{B_{5p}}} - {k} {p_{B_{1p}}}$ to construct the optimal estimator to Alice's corresponding variables $x_{A_x}$ and $p_{A_p}$, where $k = \sqrt {\eta g{T_A}{T_1}{T_2}{{\left( {{V_B} - 1} \right)} \mathord{\left/
 {\vphantom {{\left( {{V_B} - 1} \right)} {\left( {{V_B} + 1} \right)}}} \right.
 \kern-\nulldelimiterspace} {\left( {{V_B} + 1} \right)}}}$ for heterodyne detection to reduce the interference from ${\hat B_2}$, specifically minimizing the variance of ${x_{{B_x}}}$. We can use the same method to calculate the classical mutual information between Alice and Bob. The information Eve has is again given by Eq. (\ref{eq}) and the first part of it remains unchanged. But for the second part of it, we need to add modes $I_0$ and J to represent the inherent noise of the amplifier in this case. Then ${\chi _{BE}}$ is calculated from the following equations
\begin{eqnarray}
 {\chi _{BE}} &=& S({\rho _{{A_1}{A_3}{B_3}{B_1}}}) - S\Big(\rho _{{A_1}{A_3}IJFG{B_6}{B_7}}^{{x_{{B_x}}},{p_{{B_p}}}}\Big) \nonumber \\
 \quad \;\; &=& \sum\limits_{i = 1}^4 {G\Big(\frac{{{\lambda _i} - 1}}{2}\Big) - } \sum\limits_{i = 5}^{12} {G\Big(\frac{{{\lambda _i} - 1}}{2}\Big)}.
\end{eqnarray}

Therefore it is necessary to derive the covariance matrix $\gamma _{{A_1}{A_3}IJFG{B_6}{B_7}}^{{x_{B_x}},{p_{B_p}}}$ , which we can get after measuring the mode $A_1$ with heterodyne detection and measuring the modes $A_3$, $B_6$, $B_7$, $B_x$ and $B_p$ with homodyne detection as Fig.~\ref{fig1} and Fig.~\ref{fig3} (b) show.

\begin{figure}[t]
\centering
\includegraphics[width=3.2in]{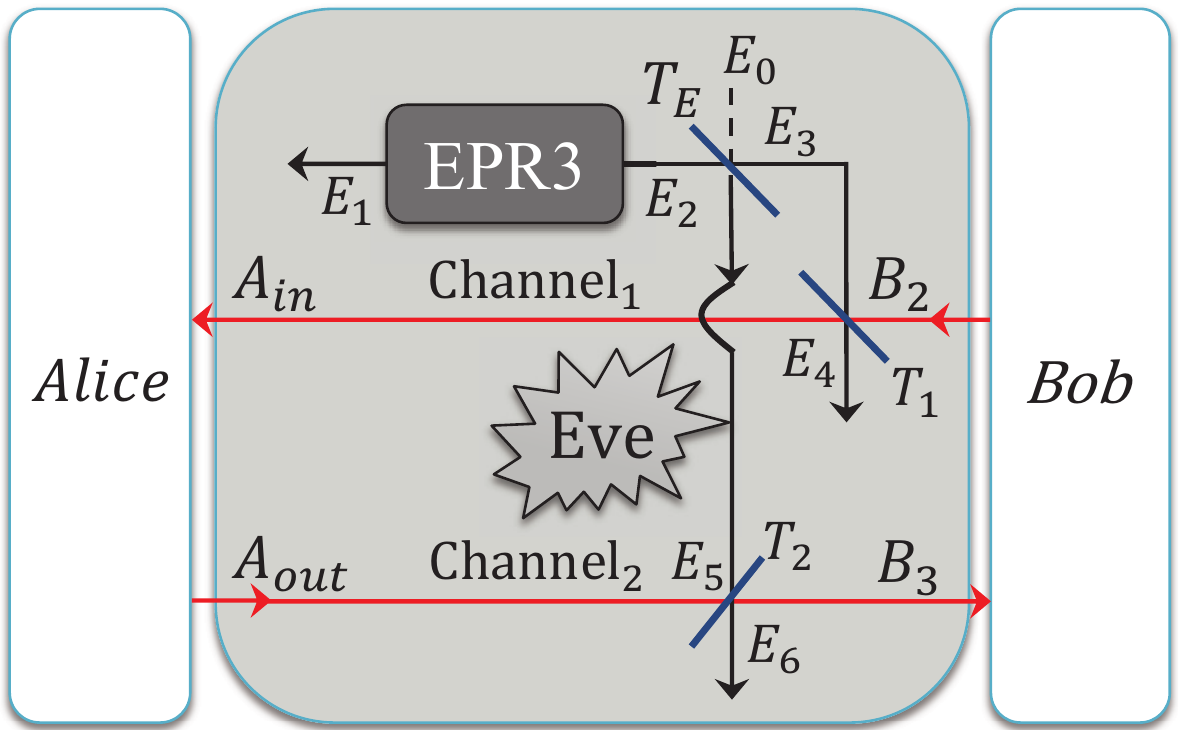}
\caption{The entanglement-based scheme of two-way CV-QKD protocols against a specific two-mode attack where Eve prepares an EPR pair (EPR3 with variance $V_{E}$), she keeps mode $E_1$ and splits mode $E_2$ with a beam splitter whose transmittance is $T_E$. $E_3$ and $E_5$ are the modes introduced into the channels. $T_1$ and $T_2$ are the channel transmission efficiencies. Here we use ${T} = {T_1} = {T_2} = {10^{ - ad/10}}$ for calculations and simulations, where $a = 0.2$~dB/km is the loss coefficient of the optical fibers, and $d$ is the length of the quantum channel.
}\label{fig5}
\end{figure}

\subsection{Simulation and discussion}
As discussed above, we need the specific description of Eve's attack to help us to obtain the final data to calculate the covariance matrices ${{\gamma _{{A_1}{A_3}{B_1}{B_3}}}}$, ${\gamma _{{A_1}{A_3}{B_{1p}}{B_6}FG}^{x_{B_x}}}$ and $\gamma _{{A_1}{A_3}IJFG{B_6}{B_7}}^{{x_{B_x}},{p_{B_p}}}$ for doing the numerical simulation. In the two-way protocol, the two-mode attack is more general and is used as Eve's attack (see Fig.~\ref{fig5}). Although the two-mode entangling cloner attack has not been proven to be the optimal attack against the two-way protocol, such an attack is the most practical benchmark to test two-way CV-QKD systems thus far in the literature~\cite{sunmaozhu_WorldScientific_2012, sunmaozhu_JPB_2012} and we again employ it here. Eve first prepares an EPR pair (EPR3 with variance $V_{E}$), she keeps mode $E_1$ and splits mode $E_2$  into $E_3$ and $E_4$ with a beam splitter whose transmittance is $T_E$. Eve then couples mode $E_3$ with the original mode $B_2$ from Bob and couples mode $E_4$ with the mode $A_{out}$ back to Bob. Therefore Eve can adjust parameter $T_E$ to reduce the interference to the mode $B_3$. When she chooses ${T_E} = {1 \mathord{\left/
 {\vphantom {1 {\left( {1 + T{T_A}} \right)}}} \right.
 \kern-\nulldelimiterspace} {\left( {1 + T{T_A}} \right)}}$ (for more details see Appendix), the mode $B_3$ can be written as

\begin{eqnarray}
{\hat B_3} = \sqrt {{T_A}} T{\hat B_2} + \sqrt {\left( {1 - {T_A}} \right)T} {\hat A_2}
~ +\sqrt {\left( {1 - T} \right)\left( {1 + T{T_A}} \right)} {\hat E_0},
\end{eqnarray}
where $E_0$ represents the vacuum state. Thus, we can get the covariance matrices ${{\gamma _{{A_1}{A_3}{B_1}{B_3}}}}$, ${\gamma _{{A_1}{A_3}{B_{1p}}{B_6}FG}^{x_{B_x}}}$ and $\gamma _{{A_1}{A_3}IJFG{B_6}{B_7}}^{{x_{B_x}},{p_{B_p}}}$ for the numerical simulation (see the Appendix for a detailed calculation).

The parameters affecting the value of the secret key rate are the reconciliation efficiency $\beta$, the variance of Alice's and Bob's modulation: $(V_A-1)$ and $(V_B-1)$, the transmittance of the beamsplitter at Alice's side $T_A$, the transmission efficiency $T$, the efficiency $\eta $ and the electronic noise ${\upsilon _{el}}$ of the detector. The parameters ${V_A}$, ${V_B}$, $\beta$, $\eta $ and ${\upsilon _{el}}$ are fixed in all simulations. The variance $V_A = V_B = 40$ which allows for the reconciliation efficiency of $\beta=0.948$, $\varepsilon = 0.02$, $\eta  = 0.552$ , and ${\upsilon _{el}} = 0.015$, which are standard in one-way CV-QKD experiments \cite{Jouguet_nature_2013}. We choose $T_A=0.4$ as the value of the beamsplitter transmittance at Alice's side and different levels of channel noise $\varepsilon = 0.005, 0.02, 0.2$.

\begin{figure}[t]
\includegraphics[width=6.2in]{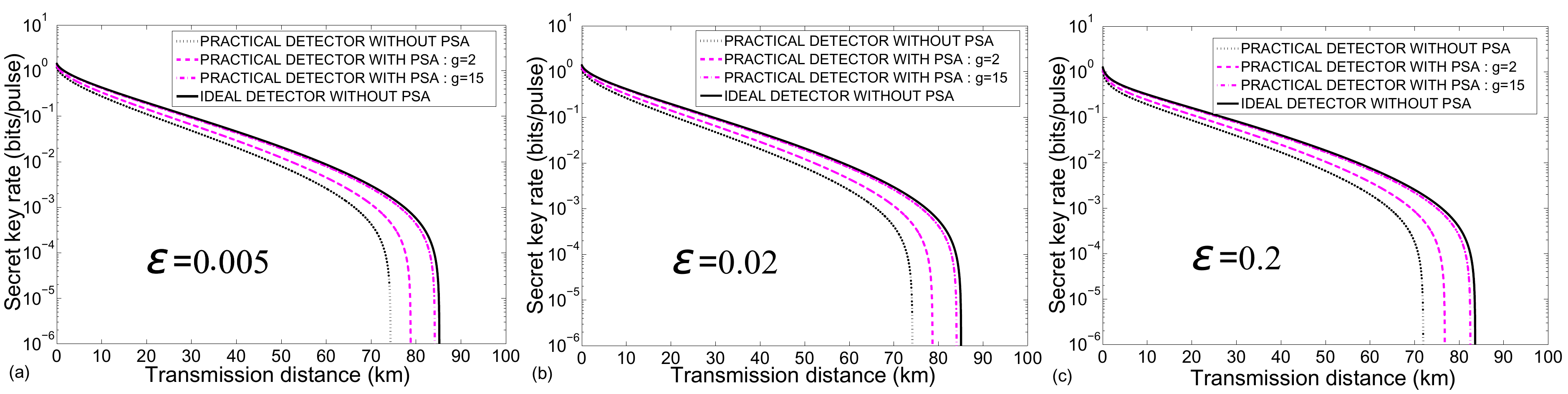}
\centering
\caption{ A comparison among the secret key rates under the following situations: no amplifier ($g=1$), using a phase sensitive amplifier whose gain is 2 or 15 with an imperfect homodyne detector ($\eta  = 0.552,{\upsilon _{el}} = 0.015$), and no amplifier with a perfect homodyne detector ($\eta  = 1$) under different levels of channel noise: (a)$\varepsilon = 0.005$, (b)$\varepsilon = 0.02$, (c)$\varepsilon = 0.2$. The reconciliation efficiency $\beta$ is 0.948~\cite{Jouguet_nature_2013}.
}\label{fig6}
\end{figure}
\begin{figure}[t]
\centering
\includegraphics[width=6.2in]{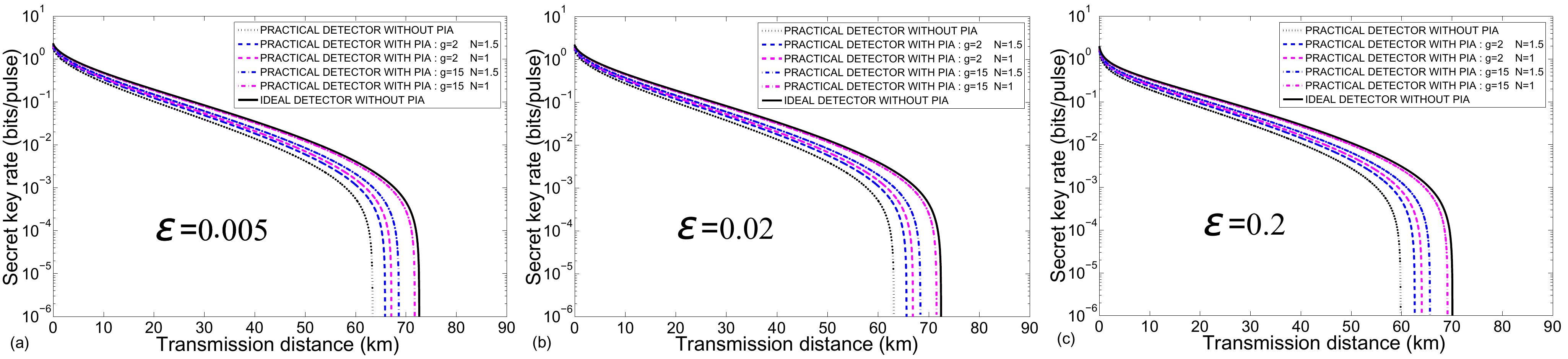}
\caption{A comparison among the secret key rates under the following situations: no amplifier ($g=1$), using a phase insensitive amplifier whose gain is $2$ or $15$ with an inherent noise of $1$ or $1.5$ with an imperfect heterodyne detector ($\eta  = 0.552, {\upsilon _{el}} = 0.015$), and no amplifier with a perfect heterodyne detector ($\eta  = 1$) under different levels of channel noise: (a)$\varepsilon = 0.005$, (b)$\varepsilon = 0.02$, (c)$\varepsilon = 0.2$. The reconciliation efficiency $\beta$ is 0.948~\cite{Jouguet_nature_2013}.
}\label{fig7}
\end{figure}

Firstly, we consider the performance of an imperfect homodyne detector with a PSA placed at the output of the quantum channel. We calculate the secret key rate $K$ as a function of distance $d$ under three situations: without using a PSA, using a PSA with gain $2$ and using a PSA with gain $15$. The simulation results are shown in Fig.~\ref{fig6}, where we find that the larger the amplification gain of the PSA, the higher the secret key rate and the longer the secure transmission distance we can achieve. We also calculate the secret key rate under perfect homodyne detection for comparison. Then we find that the new transformation of inserting a PSA at the receiver can enhance the performance of the protocol with imperfect homodyne detection under different levels of channel noise. The optimal improvement of the proposed method can approach the performances of the protocol with a perfect homodyne detector.

Secondly, we consider the performance of an imperfect heterodyne detector with a PIA placed at the output of the quantum channel. We also calculate the secret key rate under the same three situations. Additionally, we take the inherent noise of the PIA into account. The inherent noise $N$ of the PIA is set to either $1$ for minimal noise (vacuum noise) or to a more realistic value $1.5$ (referred to the input) ~\cite{Fossier_JPB_2009}. These results are shown in Fig.~\ref{fig7}. We observe that the performance of the two-way CV-QKD protocol with imperfect heterodyne detection is improved by inserting a PIA at the output of the quantum channel under different levels of channel noise. The protocol under large amplification and vacuum noise gives the highest key rate, which can approach the secret key rate of a perfect heterodyne detector. It is shown that the larger the amplification gain and the smaller the noise of PIA, the higher secret key rate and the longer secure transmission distance we can achieve. Certainly, for a practical PIA, the noise will not be as low as the vacuum noise. Therefore, in the practical system, it is important to know the tolerable PIA noise which means the most inherent noise of the PIA that the protocol can tolerate. Furthermore, the tolerable PIA noise is also the point that the method does not work.

\begin{figure}[t]
\includegraphics[width=6.1in]{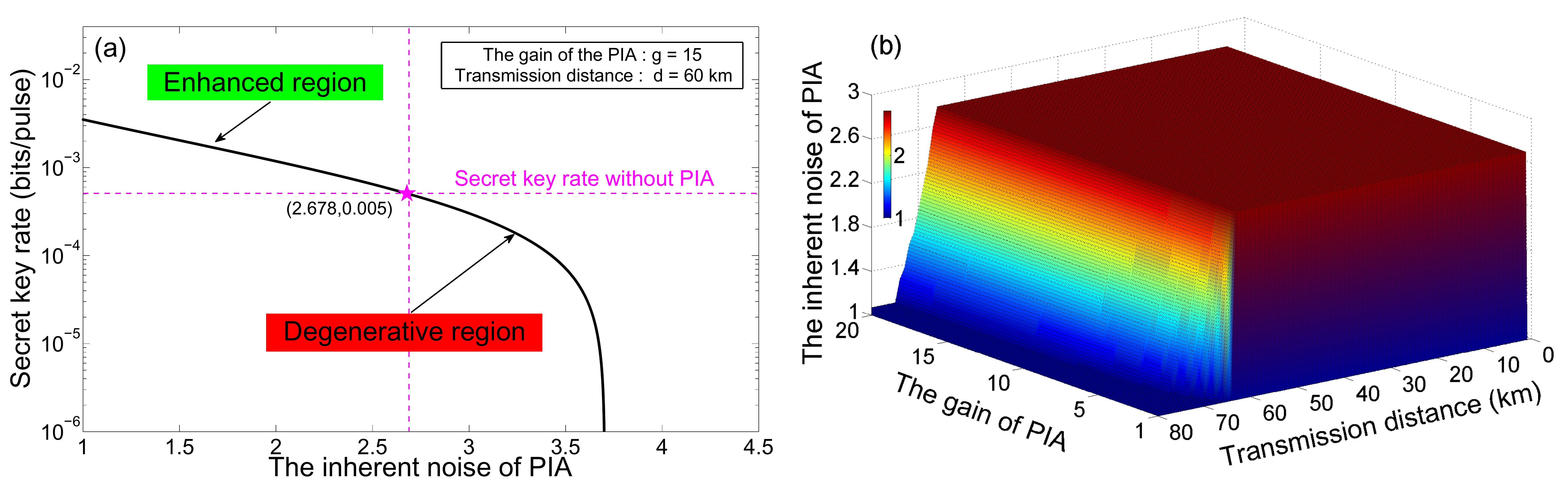}
\caption{(a) The secret key rate as a function of the inherent noise of PIA whose gain is $15$ for a fixed distance $d=60$~km and channel noise $\varepsilon = 0.02$. The horizontal reference line is drawn to represent the secret key rate of imperfect heterodyne detection without using a PIA. The region above the reference line is the enhanced region and the rest of the region is the degenerative region for the protocol. (b) The inherent noise of PIA against the gain of the PIA and the transmission distance. The tolerable PIA noise is the surface of the figure and it is constant when transmission distance is less than the maximal distance of the protocol without a PIA and decreases to 1 when the distance reaches the maximal distance of the protocol with a PIA. The whole region between the plain whose inherent noise of PIA is 1 and the tolerable PIA noise are the enhanced region.
}\label{fig8}
\end{figure}

As illustrated in Fig.~\ref{fig8} (a), we calculate the secret key rate as a function of the inherent noise of the PIA with constant gain ($g=15$) for a certain fixed distance ($d=60$~km) and channel noise ($\varepsilon = 0.02$) of the protocol. We observe that the secret key rate decreases as the noise of PIA increases, and a reference line is drawn to represent the key rate of imperfect heterodyne detection without PIA. Above the reference line, the performance of the protocol is improved by using a PIA. Therefore when the transmission distance is $60$~km, the critical value of the inherent noise of the PIA is $2.678$ which we define as tolerable PIA noise. With the method to calculate the tolerable PIA noise under a fixed gain coefficient and transmission distance, we can draw a picture of the tolerable PIA noise over the gain of the PIA and the transmission distance (see Fig.~\ref{fig8} (b)). The performance of the two-way CV-QKD protocol with imperfect heterodyne detection can be improved by placing a PIA at the output of the quantum channel, whose noise should be less than the tolerable PIA noise. From Fig.~\ref{fig8} (b), the tolerable PIA noise is constant with transmission distance less than the maximal distance the protocol without a PIA can achieve (e.g. 63.13 km for g=15). This is because the tolerable PIA noise only depends on the detector's electronic noise $\upsilon_{el}$ in this model~\cite{Fossier_JPB_2009, ZhangHeng_PhysRevA_2012}. After the maximal distance the protocol without a PIA can achieve, the tolerable PIA noise decreases and finally reaches 1 (the minimal noise of PIA), which represents the maximal distance the protocol with a PIA can achieve (e.g. 71.55 km for g=15). To improve the performance of the protocol by inserting a PIA, the inherent noise of the PIA needs to be below the tolerable amplifier noise for a fixed gain coefficient and transmission distance.

\section{\label{sec:4}Conclusion}
The imperfections of a detector can affect the performance of two-way continuous-variable quantum key distribution protocols, and are hard to adjust experimentally. In this paper, we propose a method to improve the performance of two-way continuous-variable QKD protocols by adding an adjustable optical amplifier at the output of the quantum channel, which combined with the fixed imperfect detector, can be seen as an adjustable receiver. Here we present two kinds of optical amplifier models, a phase-sensitive amplifier and a phase-insensitive amplifier, to improve the performances of homodyne detection and heterodyne detection two-way continuous-variable QKD protocols, respectively. The simulation results against a two-mode collective entangling cloner attack show that for homodyne detection, the optimal performance of adding a phase-sensitive amplifier at receiver can approach the case of using a perfect detector. On the other hand, for heterodyne detection, the optimal performance of adding a phase-insensitive amplifier at the receiver can also approach the case of using a perfect detector. We note that the proposed methods can improve the performance of protocols as long as the inherent noise of the amplifier is lower than the critical value which we define as the tolerable amplifier noise.

\section*{Acknowledgments}
This work was supported in part by the National Basic Research Program of China (973 Pro-gram) under Grant 2012CB315605 and 2014CB340102, in part by the National Science Fund for Distinguished Young Scholars of China (Grant No. 61225003), in part by the National Natural Science Foundation under Grant 61101081, 61271191, 61271193, and 61072054, and in part by the Fundamental Re-search Funds for the Central Universities. C.W. would like to thank NSERC for support.

\appendix
\section{Detailed calculation of parameter $T_E$, covariance matrices and symplectic eigenvalues}

Here we give the detailed calculation of parameter $T_E$, the covariance matrices ${{\gamma _{{A_1}{A_3}{B_1}{B_3}}}}$, ${\gamma _{{A_1}{A_3}{B_{1p}}{B_6}FG}^{x_{B_x}}}$ and $\gamma _{{A_1}{A_3}IJFG{B_6}{B_7}}^{{x_{B_x}},{p_{B_p}}}$ and their symplectic eigenvalues ${\lambda _{1 - 4}}$, ${\lambda _{5 - 10}}$ and ${\lambda _{5 - 12}}$.

The covariance matrix ${\gamma _{{A_1}{A_3}{B_1}{B_3}}}$ only depends on the system including Alice and the quantum channel whose relationships are as follows

\begin{equation}
\left\{ \begin{array}{l}
 {{\hat E}_3} = \sqrt {{T_E}} {{\hat E}_2} + \sqrt {1 - {T_E}} {{\hat E}_0} \\
 {{\hat E}_5} =  - \sqrt {1 - {T_E}} {{\hat E}_2} + \sqrt {{T_E}} {{\hat E}_0} \\
 {{\hat A}_{in}} = \sqrt {{T_1}} {{\hat B}_2} + \sqrt {1 - {T_1}} {{\hat E}_3} \\
 {{\hat E}_4} =  - \sqrt {1 - {T_1}} {{\hat B}_2} + \sqrt {{T_1}} {{\hat E}_3} \\
 {{\hat A}_{out}} = \sqrt {{T_A}} {{\hat A}_{in}} + \sqrt {1 - {T_A}} {{\hat A}_2} \\
 {{\hat A}_3} =  - \sqrt {1 - {T_A}} {{\hat A}_{in}} + \sqrt {{T_A}} {{\hat A}_2} \\
 {{\hat B}_3} = \sqrt {{T_2}} {{\hat A}_{out}} + \sqrt {1 - {T_2}} {{\hat E}_5} \\
 {{\hat E}_6} =  - \sqrt {1 - {T_2}} {{\hat A}_{out}} + \sqrt {{T_2}} {{\hat E}_5} \\
 \end{array} \right.,
\end{equation}
where $B_2$, $B_3$ and $E_0$ represents the original mode from Bob, the mode back to Bob and the vacuum state. $T_1$ and $T_2$ are the channel transmission efficiency, which we use ${T} = {T_1} = {T_2}$ for calculation here. $T_A$ and $T_E$ are the transmittances in Alice and Eve. After simplification, the state $B_3$ is

\begin{eqnarray}
{\hat B_3} &=& \sqrt {{T_A}} T{\hat B_2} + \sqrt {\left( {1 - {T_A}} \right)T} {\hat A_2} \\
\nonumber&~& + \left( {\sqrt {{T_A}{T_E}\left( {1 - {T_1}} \right){T_2}}  - \sqrt {\left( {1 - {T_2}} \right)\left( {1 - {T_E}} \right)} } \right){\hat E_2} \\
\nonumber&~& + \left( {\sqrt {{T_A}\left( {1 - {T_E}} \right)\left( {1 - {T_1}} \right){T_2}}  - \sqrt {{T_E}\left( {1 - {T_2}} \right)} } \right){\hat E_0}.
\end{eqnarray}

Therefore we can adjust parameter $T_E$ to reduce the interference of the mode $B_3$. When ${T_E} = {1 \mathord{\left/
 {\vphantom {1 {\left( {1 + T{T_A}} \right)}}} \right.
 \kern-\nulldelimiterspace} {\left( {1 + T{T_A}} \right)}}$, $B_3$ becomes

\begin{equation}
{\hat B_3} = \sqrt {{T_A}} T{\hat B_2} + \sqrt {\left( {1 - {T_A}} \right)T} {\hat A_2} + \sqrt {\left( {1 - T} \right)\left( {1 + T{T_A}} \right)} {\hat E_0}.
\end{equation}

The covariance matrix ${{\gamma _{{A_1}{A_3}{B_1}{B_3}}}}$ becomes

\begin{eqnarray}
\left( {\begin{array}{*{20}{c}}
   {{V_A}\cdot{{\rm{I}}_2}} & {\sqrt {{T_A}(V_A^2 - 1)} \cdot{\sigma _z}} & {0\cdot{{\rm{I}}_2}} & {{\gamma _{{A_1}{B_3}}}\cdot{\sigma _z}}  \\
   {\sqrt {{T_A}(V_A^2 - 1)} \cdot{\sigma _z}} & {{\gamma _{{A_3}}}\cdot{{\rm{I}}_2}} & {{\gamma _{{A_3}{B_1}}}\cdot{\sigma _z}} & {{\gamma _{{A_3}{B_3}}}\cdot{{\rm{I}}_2}}  \\
   {0\cdot{{\rm{I}}_2}} & {{\gamma _{{A_3}{B_1}}}\cdot{\sigma _z}} & {{V_B}\cdot{{\rm{I}}_2}} & {{\gamma _{{B_1}{B_3}}}\cdot{\sigma _z}}  \\
   {{\gamma _{{A_1}{B_3}}}\cdot{\sigma _z}} & {{\gamma _{{A_3}{B_3}}}\cdot{{\rm{I}}_2}} & {{\gamma _{{B_1}{B_3}}}\cdot{\sigma _z}} & {{\gamma _{{B_3}}}\cdot{{\rm{I}}_2}}  \\
\end{array}} \right),
\end{eqnarray}
where ${{{\rm I}_n}}$ is the $n \times n$ identity matrix and ${\sigma _z}$ = diag (1, -1), $V_B$, $V_A$ and $V_E$ are the variance of EPR1, EPR2 and EPR3. Here we choose ${V_E} = 1 + {{2T\varepsilon } \mathord{\left/
 {\vphantom {{2T\varepsilon } {\left( {1 - T} \right)}}} \right.
 \kern-\nulldelimiterspace} {\left( {1 - T} \right)}}$ for keeping the average excess noise of forward and backward path as $\varepsilon$. We use the average excess noise $\varepsilon = {{\left( {{\varepsilon _1}+ {\varepsilon _2}} \right)} \mathord{\left/
 {\vphantom {{\left( {{\varepsilon _1} + {\varepsilon _2}} \right)} 2}} \right.
 \kern-\nulldelimiterspace} 2}$ as the average value of forward path excess noise $\varepsilon_1$ and backward path excess noise $\varepsilon_2$. $\gamma _{{A_3}{B_3}}$, $\gamma _{A_3}$, $\gamma _{{A_3}{B_1}}$, $\gamma _{{A_3}{B_3}}$, $\gamma _{B_3}$ and $\gamma _{{B_1}{B_3}}$ are

\begin{equation}
\left\{ \begin{array}{l}
 {\gamma _{{A_1}{B_3}}} = \sqrt {T(1 - {T_A})(V_A^2 - 1)}  \\
 {\gamma _{{A_3}}}\;\;\; = {T_A}{V_A} + T(1 - {T_A}){V_B} + \frac{{T\left( {1 - {T_A}} \right){T_A}}}{{{{\left( {1 + T{T_A}} \right)}^2}}}{V_E} + \frac{{T\left( {1 - T} \right){T_A}\left( {1 - {T_A}} \right)}}{{1 + T{T_A}}} \\
 {\gamma _{{A_3}{B_1}}} =  - \sqrt {T(1 - {T_A})(V_B^2 - 1)}  \\
 {\gamma _{{A_3}{B_3}}} = \sqrt {T{T_A}(1 - {T_A})} \left( {{V_A} - T{V_B} - 1 + T} \right) \\
 {\gamma _{{B_3}}}\;\;\; = {T_A}{T^2}{V_B} + (1 - {T_A})T{V_A} + \left( {1 - T} \right)\left( {1 + T{T_A}} \right) \\
 {\gamma _{{B_1}{B_3}}} = T\sqrt {{T_A}(V_B^2 - 1)}  \\
 \end{array} \right..
\end{equation}

Similarly, the covariance matrix $\gamma _{{A_1}{A_3}GF{B_6}{B_{1p}}}^{{x_{B_x}}}$ can be written as

\begin{eqnarray}
\gamma _{{A_1}{A_3}GF{B_6}{B_{1p}}}^{{x_{{B_x}}}} &=& {\gamma _{{A_1}{A_3}GF{B_6}{B_{1p}}}} \\
\nonumber &-& \sigma _{{A_1}{A_3}GF{B_6}{B_{1p}}{B_x}}^T \cdot {\left( {X{\gamma _{{B_x}}}X} \right)^{-1}} \cdot {\sigma _{{A_1}{A_3}GF{B_6}{B_{1p}}{B_x}}}.
\end{eqnarray}
where $X$ = diag(1, 0) and the inverse is a pseudo inverse. The matrices ${\gamma _{{A_1}{A_3}GF{B_6}{B_{1p}}}}$, ${\gamma _{{B_x}}}$ and ${\sigma _{{A_1}{A_3}GF{B_6}{B_{1p}}{B_x}}}$ can all be derived from the decomposition of the matrix

\begin{equation}
{\gamma _{{A_1}{A_3}GF{B_6}{B_{1p}}{B_x}}} = \left[ {\begin{array}{*{20}{c}}
   {{\gamma _{{A_1}{A_3}GF{B_6}{B_{1p}}}}} & {\sigma _{{A_1}{A_3}GF{B_6}{B_{1p}}{B_x}}^T}  \\
   {{\sigma _{{A_1}{A_3}GF{B_6}{B_{1p}}{B_x}}}} & {{\gamma _{{B_x}}}}  \\
\end{array}} \right].
\end{equation}

The above matrix can be derived with appropriate rearrangement of lines and columns from the matrix describing the system (see Fig.~\ref{fig3} (a))
\begin{eqnarray}
\nonumber{\gamma _{{A_1}{A_3}{B_{1p}}{B_6}{B_x}FG}} &=& \Gamma _{{B_{1x}}{B_5}}^X\cdot Y_{{B_3}{F_0}}^{BS}\cdot Y_{{B_3}}^{PSA}\cdot({\gamma _{{A_1}{A_3}{B_{1p}}{B_{1x}}{B_3}}} \oplus {\gamma _{{F_0}G}})\\
&~&\cdot{\left( {Y_{{B_3}}^{PSA}} \right)^T}\cdot{\left( {Y_{{B_3}{F_0}}^{BS}} \right)^T}\cdot{\left( {\Gamma _{{B_{1x}}{B_5}}^X} \right)^T},
\end{eqnarray}
where ${\gamma _{{A_1}{A_3}{B_{1p}}{B_{1x}}{B_3}}}$ can be got from ${\gamma _{{A_1}{A_3}{B_3}{B_1}}}$ through a beam splitter which transmittance is $0.5$, while ${\gamma _{{F_0}G}}$ is the matrix that describes the EPR of variance ${\upsilon}$ used to model the detector¡¯s electronic noise. ${\upsilon}$ takes the appropriate value for the homodyne or heterodyne detection case (Sec.~\ref{sec:2}).

Then the matrices $Y_{{B_3}}^{PSA} = {{\rm I}_8} \oplus {Y^{PSA}} \oplus {{\rm I}_4}$ and $Y_{{B_3}{F_0}}^{BS}$ describes the beam splitter transformation that models the inefficiency of the detector and acts on modes $B_3$ and $F_0$. It is given by the expression
\begin{equation}
{Y^{BS}} = \left[ {\begin{array}{*{20}{c}}
   {\sqrt \eta  \cdot{{\rm{I}}_2}} & {\sqrt {1 - \eta } \cdot{{\rm{I}}_2}}  \\
   { - \sqrt {1 - \eta } \cdot{{\rm{I}}_2}} & {\sqrt \eta  \cdot{{\rm{I}}_2}}  \\
\end{array}} \right].
\end{equation}
\begin{equation}
Y_{{B_3}{F_0}}^{BS} = {{\rm{I}}_8} \oplus {Y^{BS}} \oplus {{\rm{I}}_2},
\end{equation}

The matrix ${\Gamma _x}$ is the CNOT gate\cite{G_Patron_PhD_2007,Yoshikawa_PhysRevLett_2008,Nielsen_QCQI} that transfers $B_5$ and $B_{1x}$ into modes $B_6$ and $B_x$. It is given by the expression

\begin{equation}
\Gamma _{{B_{1x}}{B_5}}^X = {{\rm{I}}_6} \oplus {\Gamma _x} \oplus {{\rm{I}}_4}\;,{\rm{with}}\;{\Gamma _x} = \left[ {\begin{array}{*{20}{c}}
   1 & 0 & { - k} & 0  \\
   0 & 1 & 0 & 0  \\
   0 & 0 & 1 & 0  \\
   0 & k & 0 & 1  \\
\end{array}} \right].
\end{equation}

We now have all the elements required to proceed to the calculation of the covariance matrix ${{\gamma _{{A_1}{A_3}{B_1}{B_3}}}}$. Then the covariance matrix $\gamma _{{A_1}{A_3}IJFG{B_6}{B_7}}^{{x_{B_x}},{p_{B_p}}}$ can be calculated in the same method with two CNOT gate\cite{G_Patron_PhD_2007,Yoshikawa_PhysRevLett_2008,Nielsen_QCQI} ${\Gamma _x}$ and ${\Gamma _p}$, which transfer $B_{5x}$, $B_{1x}$ into modes $B_6$, $B_x$ and $B_{5p}$, $B_{1p}$ into modes $B_7$, $B_p$

\begin{eqnarray}
\left[ \begin{array}{l}
 {B_x} \\
 {B_6} \\
 \end{array} \right]{\rm{ = }}{\Gamma _x}\left[ \begin{array}{l}
 {B_{5x}} \\
 {B_{1x}} \\
 \end{array} \right],\;{\rm{with}}\;{\Gamma _x}{\rm{ = }}\left[ {\begin{array}{*{20}{c}}
   1 & 0 & { - k} & 0  \\
   0 & 1 & 0 & 0  \\
   0 & 0 & 1 & 0  \\
   0 & k & 0 & 1  \\
\end{array}} \right].
\end{eqnarray}

\begin{eqnarray}
\left[ \begin{array}{l}
 {B_p} \\
 {B_7} \\
 \end{array} \right]{\rm{ = }}{\Gamma _p}\left[ \begin{array}{l}
 {B_{5p}} \\
 {B_{1p}} \\
 \end{array} \right],\;{\rm{with}}\;{\Gamma _p}{\rm{ = }}\left[ {\begin{array}{*{20}{c}}
   1 & 0 & 0 & 0  \\
   0 & 1 & 0 & k  \\
   { - k} & 0 & 1 & 0  \\
   0 & 0 & 0 & 1  \\
\end{array}} \right].
\end{eqnarray}

Finally, we need to calculate the symplectic eigenvalues of the covariance matrices ${{\gamma _{{A_1}{A_3}{B_1}{B_3}}}}$, ${\gamma _{{A_1}{A_3}{B_{1p}}{B_6}FG}^{x_{B_x}}}$ and $\gamma _{{A_1}{A_3}IJFG{B_6}{B_7}}^{{x_{B_x}},{p_{B_p}}}$. Given an arbitrary $N$-mode covariance matrix $\gamma$, there exists a symplectic matrix $S$ such that

\begin{equation}
\gamma  = S{\gamma ^ \oplus }{S^T},\quad {\gamma ^ \oplus } = \mathop  \oplus \limits_{k = 1}^N {\lambda _k} \cdot {{\rm I}_2},
\end{equation}
where the diagonal matrix ${\gamma ^ \oplus }$ is called the Williamson form of $\gamma$, and the $N$ positive quantities $\lambda _k$ are called the symplectic eigenvalues of $\gamma$ \cite{Weedbrook_RevModPhys2012}. Here the symplectic spectrum $\left\{ {{\lambda _k}} \right\}_{k = 1}^N$ can be easily computed as the standard eigenspectrum of the matrix $\left| {i\Omega \gamma } \right|$ \cite{Weedbrook_RevModPhys2012}, where the modulus must be understood in the operational sense. Here $\Omega$ is the symplectic form

\begin{equation}
\Omega  = \mathop  \oplus \limits_{k = 1}^N {\kern 1pt} {\kern 1pt} \left[ {\begin{array}{*{20}{c}}
   0 & 1  \\
   { - 1} & 0  \\
\end{array}} \right].
\end{equation}


\section*{References}
\begin{thebibliography}{10}

\bibitem{Gisin_RevModPhys_2002}
Gisin~N, Ribordy~G, Tittel~W and Zbinden~H 2002 \RMP \textbf{74} 145

\bibitem{Scarani_RevModPhys_2009}
Scarani~V, Bechmann-Pasquinucci~H, Cerf~N~J, Du\v{s}ek~M, L\"utkenhaus~N and Peev~M 2009 \RMP \textbf{81} 1301

\bibitem{Weedbrook_RevModPhys2012}
Weedbrook~C, Pirandola~S, Garc\'ia-Patr\'on~R, Cerf~N~J, Ralph~T~C, Shapiro~J~H and Lloyd~S 2012 \RMP \textbf{84} 621

\bibitem{Jouguet_nature_2013}
Jouguet~P, Kunz-Jacques~S, Leverrier~A, Grangier~P and Diamanti~E 2013 \emph{Nat.Photon.} \textbf{7} 378

\bibitem{Grosshans_PhysRevLett_2002}
Grosshans~F and Grangier~P 2002 \PRL \textbf{88} 057902

\bibitem{Weedbrook_PhysRevLett_2004}
Weedbrook~C, Lance~A~M, Bowen~W~P, Symul~T, Ralph~T~C and Lam~P~K 2004 \PRL \textbf{93} 170504

\bibitem{grosshans_nature_2003}
Grosshans~F, Van Assche~G, Wenger~J, Brouri~R, Cerf~N~J and Grangier~P 2003 \emph{Nature} \textbf{421} 238

\bibitem{Lance_PRL_2005}
Lance~A~M, Symul~T, Sharma~V, Weedbrook~C, Ralph~T~C and Lam~P~K 2005 \PRL \textbf{95} 180503

\bibitem{Lodewyck_PhysRevA_2007}
Lodewyck~J, Bloch~M, Garc\'ia-Patr\'on~R, Fossier~S, Karpov~E, Diamanti~E, Debuisschert~T, Cerf~N~J, Tualle-Brouri~R, McLaughlin~S~W and Grangier~P 2007 \emph{Phys. Rev. A} \textbf{76} 042305

\bibitem{Grosshans_PhysRevLett_2005}
Grosshans~F 2005 \PRL \textbf{94} 020504

\bibitem{Navascues_PhysRevLett_2005}
Navascu\'{e}s~M and Ac\'{i}n~A 2005 \PRL \textbf{94} 020505

\bibitem{Renner_PhysRevLett_2009}
Renner~R and Cirac~J~I 2009 \PRL \textbf{102} 110504

\bibitem{Furrer_PhysRevLett_2012}
Furrer~F, Franz~T, Berta~M, Leverrier~A, Scholz~V~B, Tomamichel~M and Werner~R~F 2012 \PRL \textbf{109} 100502

\bibitem{Leverrier_PhysRevLett_2013}
Leverrier~A, Garc\'ia-Patr\'on~R, Renner~R and Cerf~N~J 2013 \PRL \textbf{110} 030502

\bibitem{pirandola_nature_2008}
Pirandola~S, Mancini~S, Lloyd~S and Braunstein~S~L 2008 \emph{Nat. Phys.} \textbf{4} 726

\bibitem{sunmaozhu_WorldScientific_2012}
Sun~M, Peng~X, Shen~Y and Guo~H \emph{Int. J. Quantum Inform.} 2012 \textbf{10} 1250059

\bibitem{Fossier_JPB_2009}
Fossier~S, Diamanti~E, Debuisschert~T, Tualle-Brouri~R and Grangier~P 2009 \JPB \textbf{42} 114014

\bibitem{ZhangHeng_PhysRevA_2012}
Zhang~H, Fang~J and He~G 2012 \emph{Phys. Rev. A} \textbf{86} 022338

\bibitem{sunmaozhu_JPB_2012}
Sun~M, Peng~X and Guo~H 2013 \JPB \textbf{46} 085501

\bibitem{Caves_PhysRevD_1982}
Caves~C~M 1982 \emph{Phys. Rev. D} \textbf{26} 1817

\bibitem{G_Patron_PhD_2007}
Garc\'{i}a-Patr\'{o}n~R 2007 \emph{Ph.D. thesis} ULB Bruxelles

\bibitem{Yoshikawa_PhysRevLett_2008}
Yoshikawa~J~I, Miwa~Y, Huck~A, Andersen~U~L, van Loock~P and Furusawa~A 2008 \PRL \textbf{101} 250501

\bibitem{Nielsen_QCQI}
Nielsen~M~A and Chuang~I~L 2000 \emph{Quantum Computation and Quantum Communication} (Cambridge: Cambridge University Press).

\bibitem{Text}
The models we build for PSA, imperfections of homodyne detector and CNOT gate can be seen as one symplectic transformation
between ${\rho _{{A_1}{A_3}{B_1}{B_3}{B_0}{F_0}G}}$ and ${\rho _{{A_1}{A_3}{B_{1p}}{B_6}{B_x}FG}}$,
thus $S({{\rho _{{A_1}{A_3}{B_{1p}}{B_6}{B_x}FG}}}) = S({{\rho _{{A_1}{A_3}{B_1}{B_3}{B_0}{F_0}G}}}) = S({{\rho _{{A_1}{A_3}{B_1}{B_3}}}})$.

\end{thebibliography}
\end{document}